\documentstyle[twocolumn,twoside,floats,prb,aps,psfig]{revtex}

\begin{document}
\newcommand{\beq}{\begin{equation}}
\newcommand{\eeq}{\end{equation}}
\newcommand{\degree}{$^{\rm\circ} $}
\newcommand{\pcite}{\protect\cite}
\newcommand{\pref}{\protect\ref}

\title{Molecular Dynamics Studies of Sequence-directed Curvature
in Bending Locus of Trypanosome Kinetoplast DNA}

\author{Alexey K. Mazur}

\address{Laboratoire de Biochimie Th\'eorique, CNRS UPR9080\\
Institut de Biologie Physico-Chimique\\
13, rue Pierre et Marie Curie, Paris,75005, France.\\
FAX:+33[0]1.58.41.50.26. Email: alexey@ibpc.fr}

\date{\today}
\maketitle

\begin{abstract}

The macroscopic curvature induced in the double helical B-DNA by
regularly repeated adenine tracts (A-tracts) plays an exceptional role
in structural studies of DNA because this effect presents the most
well-documented example of sequence specific conformational
modulations. Recently, a new hypothesis of its physical origin has
been put forward, based upon the results of molecular dynamics
simulations of a 25-mer fragment with three A-tracts phased with the
helical screw. Its sequence, however, had never been encountered in
experimental studies, but was constructed empirically so as to
maximize the magnitude of bending in specific computational
conditions. Here we report the results of a similar investigation of
another 25-mer B-DNA fragment now with a natural base pair sequence
found in a bent locus of a minicircle DNA. It is shown that the static
curvature of a considerable magnitude and stable direction towards the
minor grooves of A-tracts emerges spontaneously in conditions
excluding any initial bias except the base pair sequence. Comparison
of the bending dynamics of these two DNA fragments reveals both
qualitative similarities and interesting differences. The results
suggest that the A-tract induced bending obtained in simulations
reproduces the natural phenomenon and validates the earlier
conclusions concerning its possible mechanism.

\end{abstract}

\section*{Introduction}

It is now generally accepted that the double helical DNA can somehow
translate its base pair sequence in tertiary structural forms. The
simplest such form is a bend. Large bends in natural DNA
were discovered nearly twenty years ago for sequences
containing regular repeats of $\rm A_nT_m,\ with\ n+m>3$, called
A-tracts \cite{Marini:82,Wu:84}. Since then this intriguing
phenomenon has been thoroughly studied, with several profound
reviews of the results published in different years
\cite{Diekmann:87a,Hagerman:90,Crothers:90,Crothers:92,Olson:96,Crothers:99}.
Every A-tract slightly deviates the helical axis towards its minor
groove, which results in significant macroscopic curvature when they
are repeated in phase with the helical screw. However, in spite of
considerable efforts spent in attempts to clarify the the physical
origin of this phenomenon it still remains a matter of debate because
all theoretical models proposed until now contradict some of the
experimental results. This problem is of general importance because
the accumulated large volume of apparently paradoxical observations on
the subject points to our lack of understanding of the fundamental
principles that govern the DNA structure.

A variety of theoretical methods based upon computer molecular
modeling have been used in order to get insight in the mechanism of
DNA bending
\cite{Zhurkin:79,Kitzing:87,Chuprina:88,Zhurkin:91,Sanghani:96}. The
most valuable are unbiased simulations where sequence dependent
effects result only from lower level interactions and are not affected
by {\em a priori} empirical knowledge of relevant structural
features \cite{Sprous:99}. Such calculations can reveal the essential
physical factors and eventually shed light upon the true mechanism of
DNA bending. We have recently reported about the first example of
stable static curvature emerging spontaneously in free molecular
dynamics simulations a B-DNA fragment with A-tracts phased with the
helical screw, and proposed a new mechanism of bending that could
explain our results as well as experimental data
\cite{Mzlanl:00,Mzlanl:00a}. However, the sequence used in these
computational experiments was artificial in the sense that it was
designed empirically so as to accelerate the development and maximize
the amplitude of bending. It was never studied experimentally,
therefore, even though it was similar to canonical DNA benders, one
could not exclude that the static bending observed in calculations was
of a different origin than that found in experiments. Here we report
the results of a similar study of a complementary 25-mer DNA duplex
$\rm AAAATGTCAAAAAATAGGCAAATTT$. This fragment is found in the center
of the first bent DNA locus investigated experimentally {\em in
vitro} \cite{Marini:82,Wu:84}. It belongs to a minicircle DNA from
the kinetoplast body of {\em Leishmania tarentolae} and, together with
several additional A-tracts, provides planar curvature that
apparently serves {\em in vivo} to facilitate the loop closure. We
have only replaced the 3'-terminal $\rm A_6$ tract in the original
fragment \cite{Wu:84} by $\rm A_3T_3$ because our preliminary
empirical observations suggested that 3'-terminal $\rm A_n$ tracts
usually need larger time to converge to characteristic conformations
with a narrow minor groove.

We show that two independent long MD trajectories starting from
straight conformations corresponding to canonical A and B-DNA forms
both converge to statically bent structures with similar bending
direction and magnitude, thus giving the first example of a natural
DNA fragment where this phenomenon is reproduced in simulations. The
results are qualitatively similar to our first report \cite{Mzlanl:00a}
as regards the kinetics of convergence and comparison with different
theoretical models of bending. At the same time, along with
convergence of the overall macroscopic DNA from, we find here a
remarkably larger than earlier degree of convergence in local backbone
and base pair conformations. These results confirm that A-tract
induced DNA bending found in calculations corresponds to the
experimental phenomenon. They provide additional information
concerning its mechanism and the intriguing relationship between the
base pair sequence and the DNA structure.

\section*{Results}

Two long MD trajectories have been computed for the complementary
duplex $\rm AAAATGTCAAAAAATAGGCAAATTT$. The first trajectory referred
to below as TJB started from the fiber canonical B-DNA structure and
continued to 10 ns. The second trajectory (TJA) started form the
fiber canonical A-DNA conformation and continued to 20 ns. The longer
duration of TJA was necessary to ensure structural convergence. The
computational protocols are detailed in Methods.

\begin{figure}
\centerline{\psfig{figure=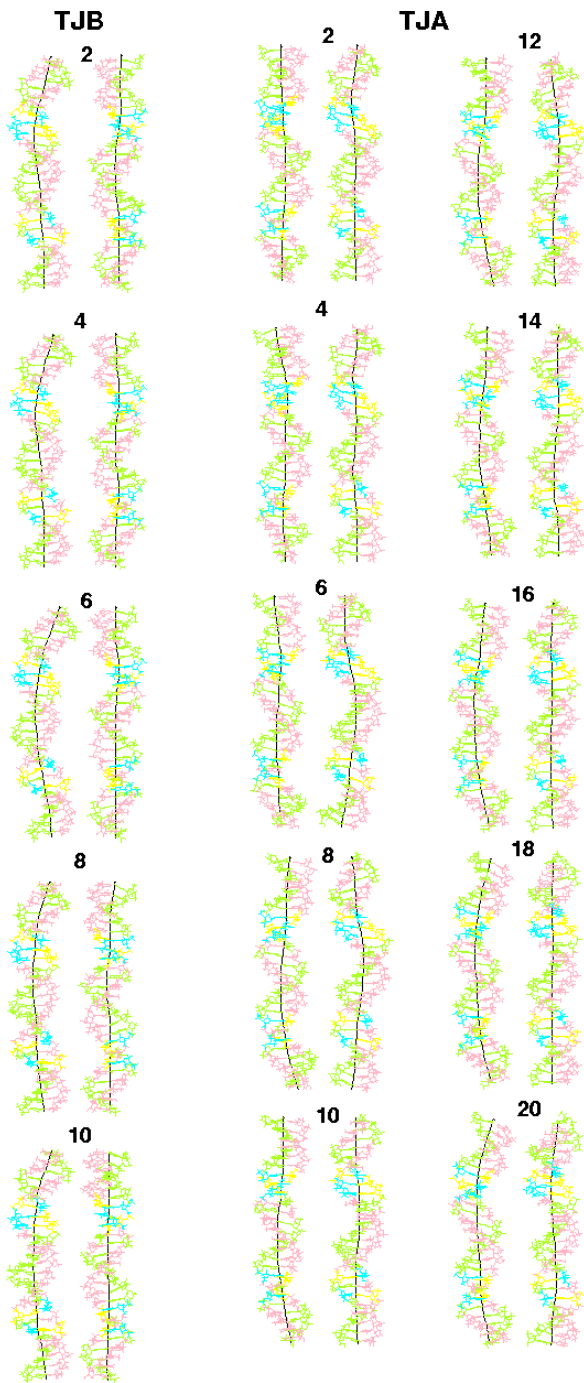,height=19cm,angle=0.,%
bbllx=-20pt,bblly=210pt,bburx=350pt,bbury=770pt,clip=t}}
\caption{\label{Ftj15av}
Consecutive average structures from TJB and TJA. The average
conformations were taken from one nanosecond intervals equally spaced
over the trajectories, namely, during the second, fourth, sixth
nanosecond, and so forth. They were all superimposed and two
perpendicular views are shown in the figure. In both trajectories,
the view is chosen according to the final orientation of the
bending plane. Namely, a straight line between the ends of the helical
axis passes through its center in the right hand projection.
 Residues are coded by colors, namely, A - green, T - red,
G - yellow, C - blue. }\end{figure}

Figure \ref{Ftj15av} shows two series of representative structures
from the two trajectories. Each structure is an average over a
nanosecond interval, with these intervals equally spaced in time. In
TJB the molecule was curved always in the same direction, but both the
shape of the helical axis and the magnitude of bending varied. The
first two structures shown in Fig. \ref{Ftj15av}a are locally bent
between the upper two A-tracts while their lower parts are nearly
straight. In contrast, in the last three structures, the planar
curvature is smoothly distributed over the entire molecule. In TJA,
distinguishable stable bending emerged only after a few nanoseconds,
but after the fourth nanosecond all average conformations were
smoothly bent. In contrast to TJB, however, the bending direction was
not stable, and by comparing the two time series of perpendicular
projections one may see that during the first 15 nanoseconds the
bending plane slowly rotated. During the final five nanoseconds the
overall bending direction was stable. In the last conformation, an
S-shaped profile of the helical axis is found in the perpendicular
projection, which indicates that there are two slightly misaligned
bends located between the three A-tracts. The orientations of the
helices in this figure was chosen separately for the two trajectories,
and one can notice that the left projection in plate (a) is close to
the right one in plate (b), in other words, the final bend directions
differed by approximately 90\degree. In TJA, the intersection of the
minor groove with the bending plane occurs close to the center of
the middle A-tract while in TJB this point is shifted towards its 3'
end. In both cases, however, the narrowed minor grooves of the
A-tracts appear at the inside edge of the curved axis.

\begin{figure}
\centerline{\psfig{figure=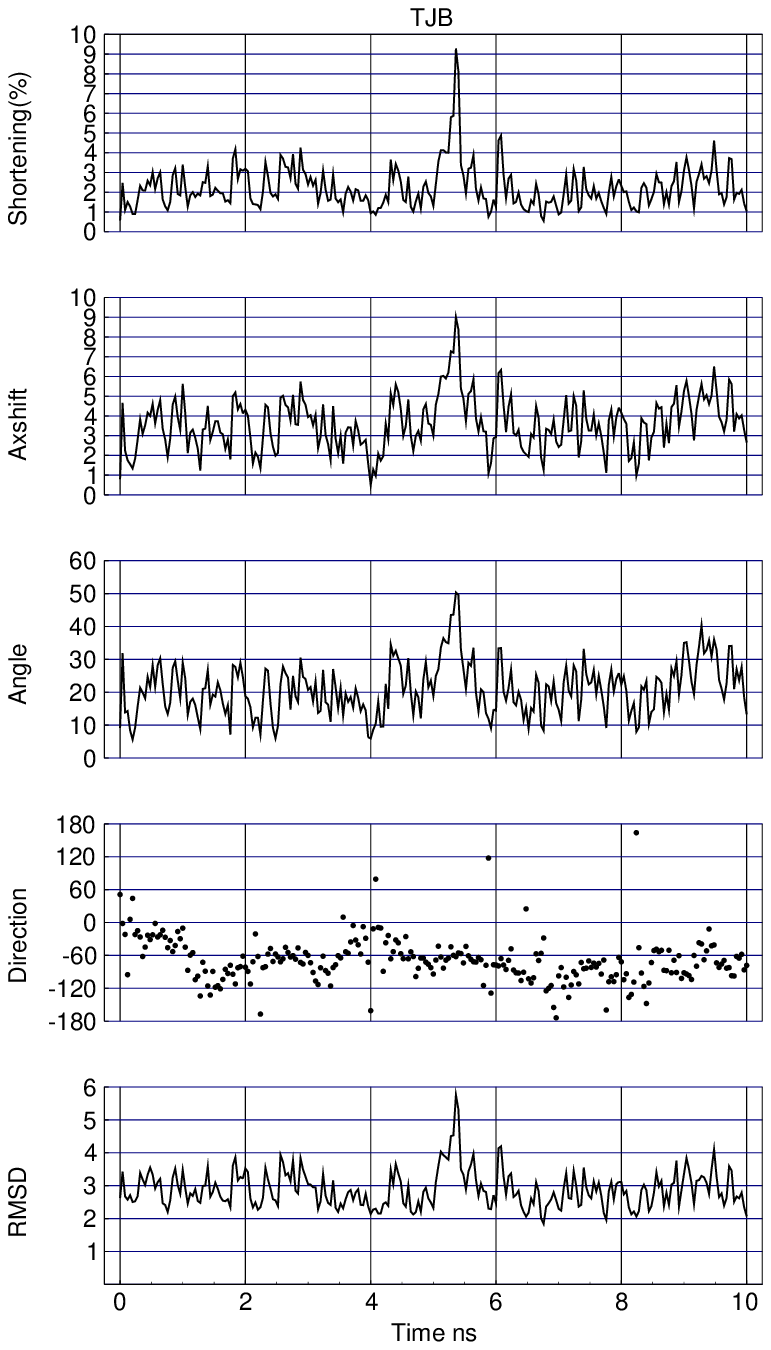,height=11cm,angle=0.,%
bbllx=60pt,bblly=-20pt,bburx=295pt,bbury=370pt,clip=t}}
\caption{(a)}
\end{figure}\addtocounter{figure}{-1}

\begin{figure}
\centerline{\psfig{figure=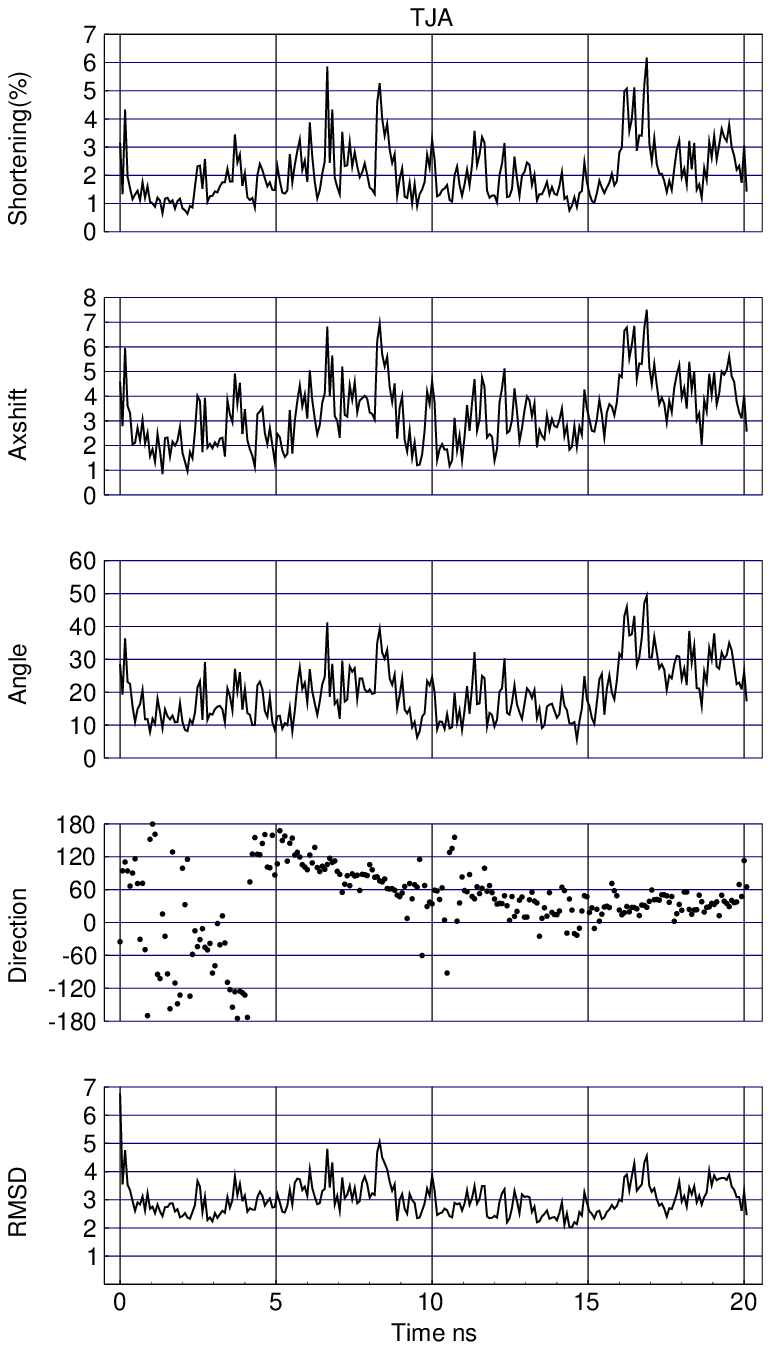,height=11cm,angle=0.,%
bbllx=60pt,bblly=-20pt,bburx=295pt,bbury=370pt,clip=t}}
\caption{(b)}
\end{figure}\addtocounter{figure}{-1}

\begin{figure}
\caption{\label{F5tc}
The time evolution of several representative structural parameters in
TJB (a) and TJA (b). Nonhydrogen atom rmsd is measured with respect to
the fiber canonical B-DNA model constructed from the published atom
coordinates \pcite{Arnott:72}. The bending direction is characterized
by the angle (given in degrees) between the plane that passes through
the two ends and the middle point of the helical axis, and the $xz$
plane of the local DNA coordinate frame at the center of the duplex.
The local frame is constructed according to the Cambridge convention
\pcite{Dickerson:89}, namely, its $x$ direction points to the major DNA
groove along the short axis of the base-pair, while the local $z$ axis
direction is adjacent to the optimal helicoidal axis. Thus, a zero
angle between the two planes corresponds to the overall bend to the
minor groove exactly at the central base pair. The bending angle is
measured between the two ends of the helical axis. The shift parameter
is the average deviation of the helical axis from the straight line
between its ends. The shortening is measured as the ratio of the
lengths of the curved axis to its end-to-end distance minus one. All
traces have been smoothed by averaging with a window of 75 ps in (a) and
150 ps in (b).}\end{figure}

Figure \ref{F5tc} displays fluctuations of various parameters that
characterize the overall bending of the double helix. In both
trajectories the rmsd from the canonical B-DNA usually fluctuates
between 2 and 4 \AA\ and correlates with the three parameters shown in
Fig. \ref{F5tc} that measure the bending magnitude. Note that in TJA
there was a short period of strong initial bending which have not been
detected in Fig. \ref{Ftj15av}. The most significant difference
between the two plates in Fig. \ref{F5tc} is in the dynamics of the
bending direction. In TJB, the final orientation of the bend was found
early and remained quite stable, although the molecule sometimes
straightened giving, simultaneously, a low bending magnitude and large
temporary fluctuations of direction. In contrast, during the first 15
nanoseconds of TJA, the bending plane made more than a half turn with
respect to the coordinate system bound to the molecule, that is a
transition occurred between oppositely bent conformations by passing
through a continuous series of bent states. Temporary transitions to
the straight state were short-living and always reversed, with the
bending resumed in the previous direction. Owing to this turn the bend
orientations in TJA and TJB converged considerably although not
exactly. The zero direction in Fig. \ref{F5tc} corresponds to bending
towards the minor groove at the fifth AT base pair of the middle
A-tract. We see that, at the end of TJA, it is shifted from zero by an
angle corresponding to rotation in one base pair step,
which gives a bend towards the minor groove close to
the center of the middle A-tract. In TJB, the direction deviates from
zero in the opposite sense by an angle corresponding to roughly two
base pair steps, resulting in a residual divergence of approximately
90\degree\ between the two trajectories. The slow kinetics of
convergence exhibited in Fig. \ref{F5tc} indicates, however, that
still better accuracy, if ever possible, would require much longer
time.

\begin{figure}
\centerline{\psfig{figure=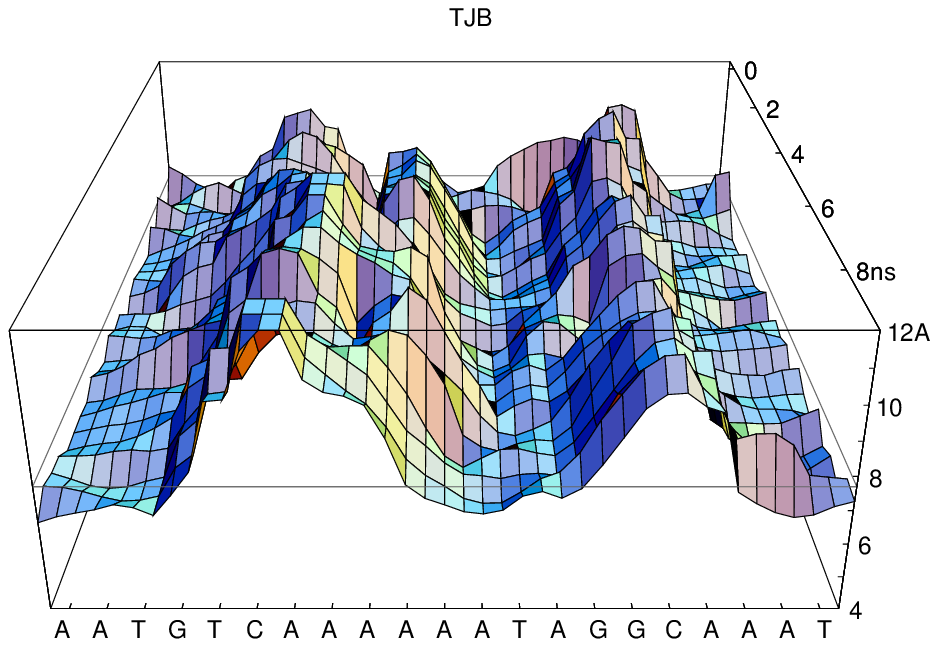,height=6cm,angle=0.}}
\caption{(a)}
\end{figure}\addtocounter{figure}{-1}

\begin{figure}
\centerline{\psfig{figure=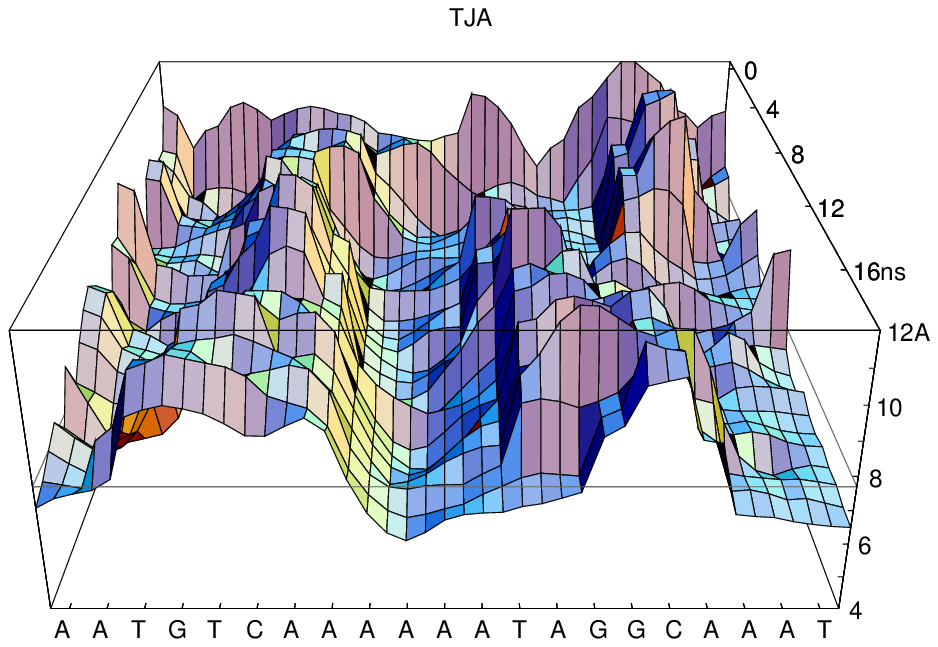,height=6cm,angle=0.}}
\caption{(b)}
\end{figure}\addtocounter{figure}{-1}

\begin{figure}
\caption{\label{Fmgkt}
The time evolution of the profile of the minor groove in TJB (a) and
TJA (b). The surfaces are formed by time-averaged successive minor
groove profiles, with that on the front face corresponding to the
final DNA conformation. The interval of averaging was 75 ps in TJB and
150 ps in TJA. The groove width is evaluated by using space traces of
C5' atoms as described elsewhere \pcite{Mzjmb:99}. Its value is given
in angstr\"oms and the corresponding canonical B-DNA level of 7.7 \AA\
is marked by the straight dotted lines on the faces of the box.}
\end{figure}

\begin{figure}
\centerline{\psfig{figure=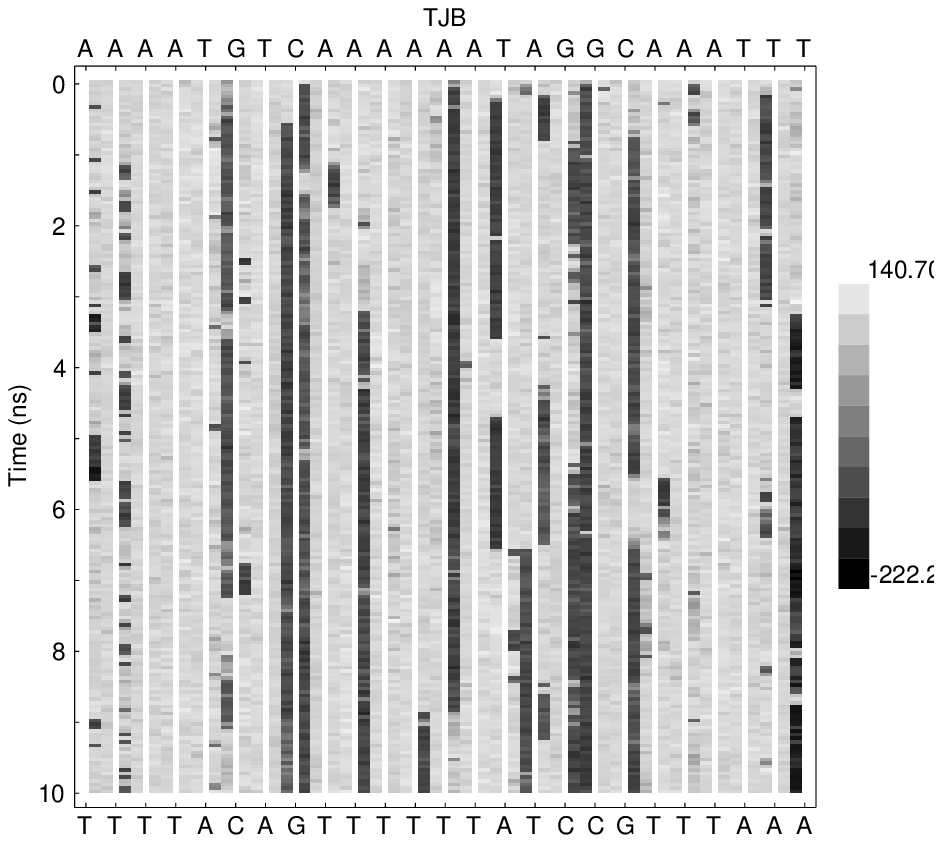,height=6cm,angle=0.}}
\caption{(a)}
\end{figure}\addtocounter{figure}{-1}

\begin{figure}
\centerline{\psfig{figure=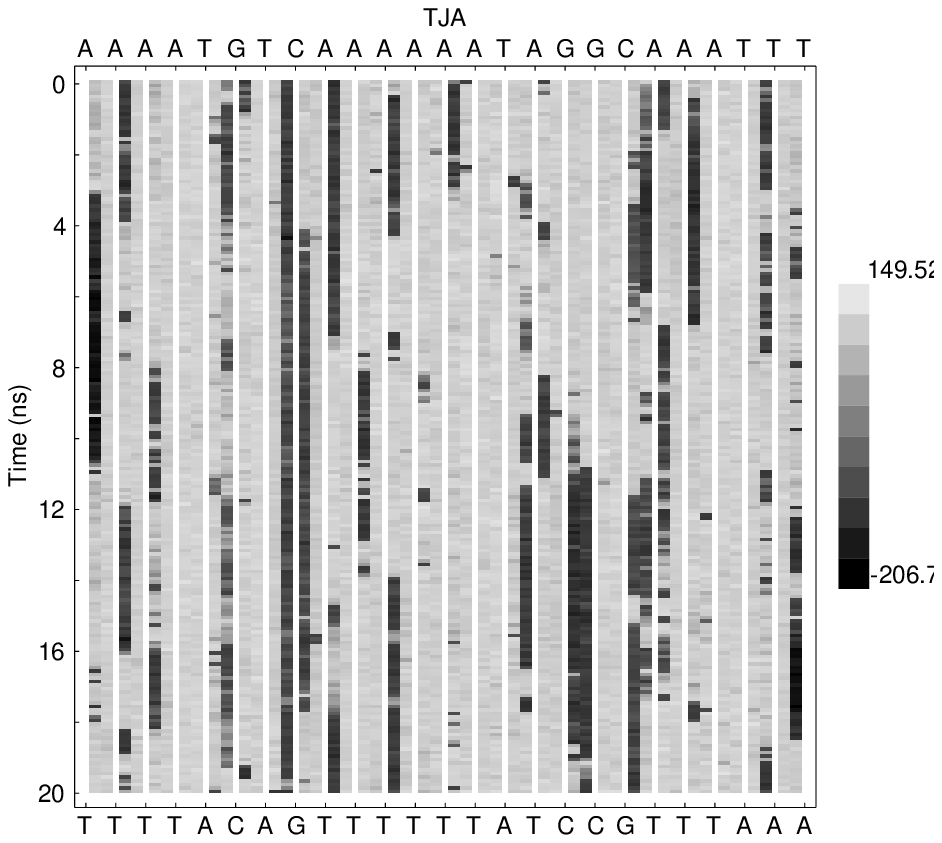,height=6cm,angle=0.}}
\caption{(b)}
\end{figure}\addtocounter{figure}{-1}

\begin{figure}
\caption{\label{FB1.B2}
Dynamics of $\rm B_I\leftrightarrow B_{II}$ transitions in TJB (a) and
TJA (b). The B$_{\rm I}$ and B$_{\rm II}$ conformations are
distinguished by the values of two consecutive backbone torsions,
$\varepsilon$ and $\zeta$. In a transition they change concertedly
from (t,g$^-$) to (g$^-$,t). The difference $\zeta -\varepsilon$ is,
therefore, positive in B$_{\rm I}$ state and negative in B$_{\rm II}$,
and it is used in as a monitoring indicator, with the corresponding
gray scale levels shown on the right in each plate. Each base pair step is
characterized by a column consisting of two sub-columns, with the left
sub-columns referring to the sequence written at the top in 5'-3'
direction from left to right. The right sub-columns refer to the
complementary sequence shown at the bottom.}\end{figure}

Figure \ref{Fmgkt} displays the time evolution of the profile of the
minor grooves in TJB and TJA. At the end of both trajectories the
minor groove width exhibits modulations phased with the helical screw.
It is significantly widened between the A-tracts and narrowed within
them by approximately 1 \AA\ with respect to the canonical level. This
magnitude of narrowing corresponds well to the values observed in
experimental structures of A-tract containing B-DNA oligomers
\cite{Dickerson:81a,Nelson:87,DiGabriele:89,Edwards:92,DiGabriele:93,Mzjmb:99}.
In TJB, the overall waving profile established early and remained more
or less constant. Interestingly, during two rather long periods, a
secondary minimum occurred close to the 5' end of the middle A-tract,
and at the same time the main central minimum sifted towards the 3'
end of this A-tract. These motions involve almost an entire helical
turn and, apparently, are concerted, which demonstrates the
possibility of medium range structural correlations along the double
helix. Comparison of Figs. \ref{Ftj15av}a and \ref{Fmgkt}a suggests
that there is no simple one-to-one relationship between bending and
minor groove modulations. Notably, the right smaller and narrower
widening corresponds to a stable and strong bending point of the
helical axis, while the left one, which is evidently larger, gives
less or no bending. In TJA, the final configuration of the minor groove
established only during the last few nanoseconds, but the final
profile has the same number and similar positions of local maxima and
minima as that in TJB. The overall minor groove dynamics in TJA looks
rather complicated and its relationship with the quasi-regular
rotation of the bending plane demonstrated in Figs. \ref{Ftj15av}b
and \ref{F5tc}b is not readily seen.

\begin{figure}
\centerline{\psfig{figure=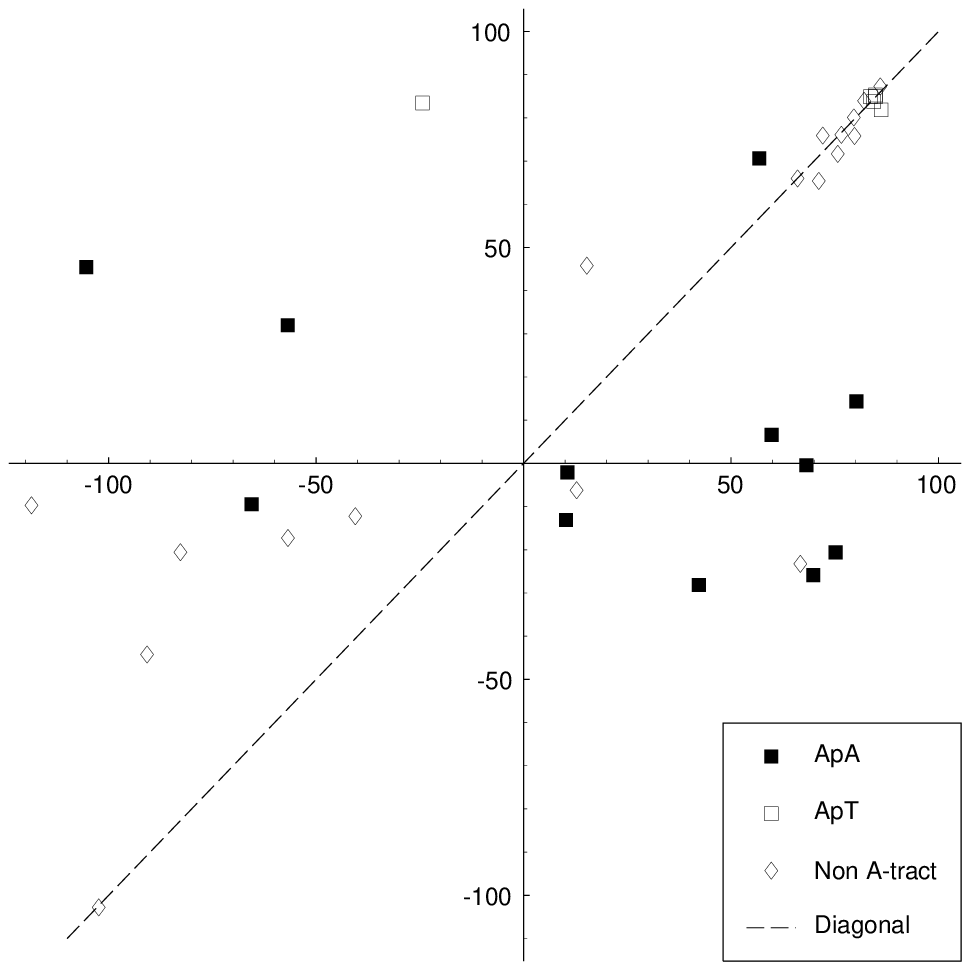,height=6cm,angle=0.,%
bbllx=0pt,bblly=0pt,bburx=290pt,bbury=290pt,clip=t}}
\caption{\label{FBBcorr}
Correlation between populations of B$_{\rm I}$ and B$_{\rm II}$
conformers in different base pair steps in TJB and TJA. Each point in
the plot represents a specific base pair step, with the corresponding
average $\zeta -\varepsilon$ values in TJB and TJA used as $x$ and $y$
coordinates, respectively. The TpT steps in A-tracts are omitted for
clarity.}\end{figure}

Figure \ref{FB1.B2} displays dynamics of $\rm B_I\leftrightarrow
B_{II}$ backbone transitions in the two trajectories. A few common
features can be noticed that have been encountered previously
\cite{Mzlanl:00,Mzlanl:00a}. For instance, in A-tracts, the B$_{\rm
II}$ conformers are commonly found in ApA steps and almost never in
TpT steps. They tend to alternate with B$_{\rm I}$ in consecutive ApA
steps. $\rm B_I\leftrightarrow B_{II}$ transitions often occur
concertedly along the same strand as well as in opposite strands. The
$\rm B_I\leftrightarrow B_{II}$ dynamics comprises all time scales
presented in these calculations and clearly involves slower motions as
well. Note, for instance, a particularly stable B$_{\rm II}$
conformer in the only GpA step available in the two strands. On the
other hand, there is some similarity in the distributions of the
B$_{\rm II}$ conformers in the two trajectories, which is a new
feature compared to our previous report \cite{Mzlanl:00,Mzlanl:00a}. It
is seen in Fig. \ref{FB1.B2} that in total ten B$_{\rm II}$
conformers were found at the end of TJB and eight in TJA. Among them
six and five, respectively, occurred in non A-tract sequences. In three
base pair steps the B$_{\rm II}$ conformers are found in both final
conformations, with all of them in non A-tract sequences. A careful
examination the two plates in Fig. \ref{FB1.B2} shows that, although
in A-tracts the preferred sites of B$_{\rm II}$ conformations differ,
in the intervening sequences their dynamics is rather similar in TJB
and TJA. This trend is demonstrated in Fig. \ref{FBBcorr} where
inter-trajectory correlations are examined for the specific base pair
step propensities to B$_{\rm I}$ and B$_{\rm II}$ conformers. We have
not included here the TpT steps in A-tracts because they strongly
prefer the B$_{\rm I}$ conformation and, therefore, are trivially
correlated. It is evident that, except for ApA steps in A-tracts,
there was certain correlation in the average populations of B$_{\rm
II}$ conformers for each specific base pair step in the two
trajectories. The ApA steps apparently can adopt both conformations
with little effect of the sequence context and the overall structure.

\begin{figure}
\centerline{\psfig{figure=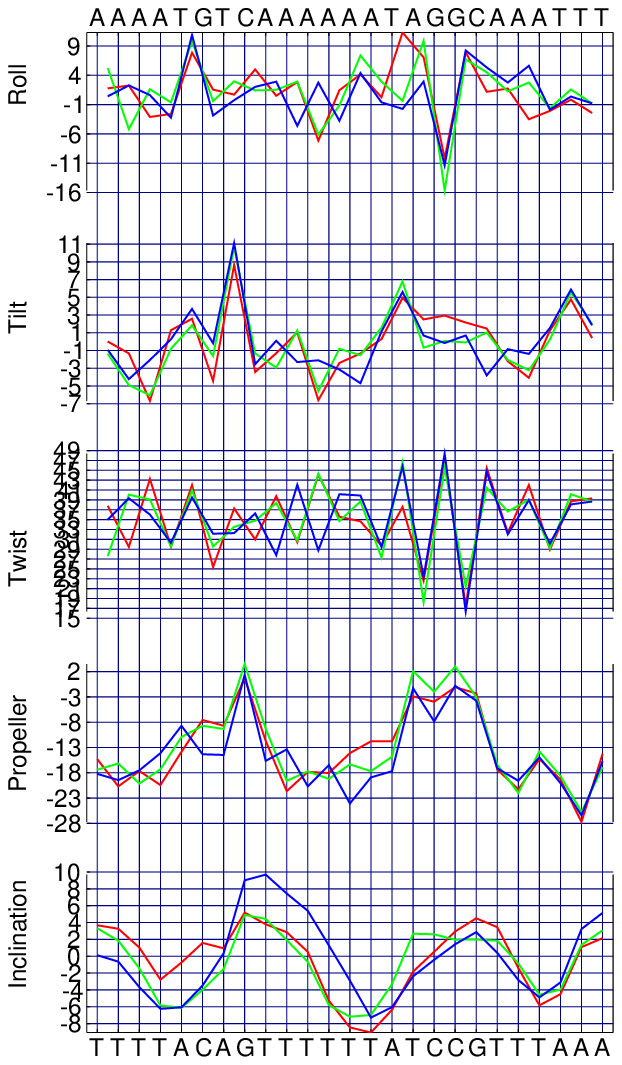,height=11cm,angle=0.,%
bbllx=55pt,bblly=-14pt,bburx=233pt,bbury=300pt,clip=t}}
\caption{\label{Fhlpa}
Sequence variations of some helicoidal parameters in representative
one nanosecond averaged structures from TJB and TJA. The sequence of
the first strand is shown on the top in 5' -- 3' direction. The
complementary sequence of the second strand is written on the bottom
in the opposite direction. All parameters were evaluated with the
Curves program \pcite{Curves:} and are given in degrees. The color
coding is: TJA 18th ns -- red, TJA 16th ns -- green and TJB 10th ns --
blue. }\end{figure}

Figure \ref{Fhlpa} shows variation of several helicoidal parameters
along the duplex in three representative one nanosecond averaged
structures. Two of them were taken from TJA, namely, the 16th and 18th
nanosecond averages which we refer to as TJA16 and TJA18. They
illustrate the scale and the character of fluctuations of these
parameters in the course of the same dynamics. The third conformation
is the last average of TJB (TJB10) and it illustrates convergence of
helical parameters in independent trajectories. We have chosen TJA16
and TJA18 because, as seen in Fig. \ref{Ftj15av}, the corresponding
two structures are particularly similar. They are both smoothly bent
in a virtually identical direction and their rmsd is only 0.95 \AA.
All parameters shown in the figure, except the inclination, exhibit
jumping alternations between consecutive base pair steps. Although they
look chaotic, there is a considerable similarity between TJA16 and
TJA18 and less significant, but still evident similarity of the two
with TJB10. Notably, a remarkable correspondence of alterations of the
twist is observed in the right-hand half of the sequence. At the same
time, even the TJA16 and TJA18 plots sometimes diverge. Note, for
instance, that the alteration in their roll traces are phased in the
central A-tract, but dephased in the other two, with a particularly
significant difference around the TpA step. These results show that,
in a statically curved double helix, the base pair helicoidal
parameters fluctuate around certain specific values that are
stable in the nanosecond time scale. There is, however, more than one
combination of these parameters for the same overall bend. At the same
time, the evident convergence of the corresponding distributions in
TJA and TJB suggests that, at least for this particular base pair
sequence, the number of such redundant combinations should not be very
large.

\subsection*{Discussion}

The computational experiments described here give the first example of
a natural DNA fragment with phased A-tracts which in free unbiased MD
simulations adopts spontaneously a statically bent shape. In the
analogous earlier calculations the static curvature has been
demonstrated for a different A-tract containing sequence constructed
artifically and never tested in experiments
\cite{Mzlanl:00,Mzlanl:00a}. The qualitative similarity between these
two simulations is evident. Trajectories starting from canonical A
and B-DNA forms both travel within the B-DNA family and, in straight,
states yield rmsd from B-DNA of around 2 \AA. TJA enters the B-DNA
family with a significant temporary bending during the first 500 ps.
Later it becomes bent in an arbitrary direction and next changes the
bend direction by slowly rotating the bending plane. This rotation
slows down after 10 ns, with the final orientation much closer to that
in TJB than the initial one. In both cases the residual discrepancy
was in the range of 60\degree\ -- 90\degree\ \cite{Mzlanl:00a}. The
final minor groove profiles are not identical, although similar for
TJA and TJB, as well as the distributions of the $\rm B_I$ and $\rm
B_{II}$ backbone conformers and base pair helicoidal parameters. The
present results, therefore, suggest that the A-tract induced
DNA bending observed in calculations here and before corresponds to
the experimental phenomenon.

At the same time, there are several clear differences. Notably, the
preferred bending direction here is closer the centers of the minor
grooves of the A-tracts, whereas the magnitude of bending is somewhat
less than in the previous calculations. The bending angle in the
average structures shown in Fig. \ref{Ftj15av} fluctuates between
12\degree\ and 25\degree\ in TJB and between 7\degree\ and 28\degree\
in TJA, with the maximal values reached at the end in both cases. The
previous artifical sequence was constructed to maximize the bending
and it showed the corresponding values beyond 35\degree\
\cite{Mzlanl:00,Mzlanl:00a}. According to the experimental estimates
made for ``good benders'' in an optimal sequence context, the
magnitude of bending is around 18\degree\ per A-tract
\cite{Crothers:99}, which in our case gives 36\degree\ for the overall
bend because the principal bending elements are the two intervening zones
between the A-tracts. The bends observed here are somewhat below this
estimate. However, in experiments, the bending magnitude differs
significantly between different A-tract sequences and depends upon
many environmental parameters that are not controlled in simulations.
One can expect to observe in calculations sequence variations of the
bending magnitude that may not exactly follow those in experiments.
Therefore, whatever the possible reasons of the apparent
discrepancy, the overall correspondence of the computed bending
magnitudes to the experimental estimates should be considered as
surprisingly good. Yet another difference is a larger than earlier
degree of similarity in the profiles of the minor groove, in the
distributions of $\rm B_{II}$ conformers, and helicoidal parameters in
trajectories starting from A and B-DNA forms. It was the most
surprising observation of our previous report that reasonably good
convergence in terms of the macroscopic bent shape of the double helix
was not accompanied by the parallel convergence of microscopic
conformational parameters. Here the two trajectories manifest clear
signs of convergence for base pair step parameters as well as for the
backbone conformations. Additional studies are necessary in order to
tell if this difference is a sequence specific property or just an
occasional effect.

In spite of these differences the results presented here support our
previous conclusion concerning the qualitative disagreement of the
computed structural dynamics of the double helix and the most popular
earlier theories of bending \cite{Mzlanl:00a}. In Figs. \ref{Fmgkt},
\ref{FB1.B2}, and \ref{Fhlpa}, multiple examples are found of strong
variability of local helical parameters in bent substates,
which argues against the local interactions and the preferred base
pair orientations as the cause of bending. All three A-tracts are
characterized by a narrowed minor groove and local minima in the
traces of the propeller twist and the inclination of base pairs.
Nevertheless, their internal structures are not homogeneous and vary
from one base pair step to another. Moreover, the structures of the
three are not the same and present another example of an ensemble of
conformational substates with a common overall shape. This pattern is
qualitatively different from that implied by the junction models of
bending \cite{Levene:83,Nadeau:89}. At the same time, the present
results are well interpreted in the framework of our model that
sees the principal cause of the bending in the backbone stiffness and
the postulated excess of its specific length over that in the
idealized regular B-DNA double helix \cite{Mzlanl:00a}. Several non
trivial observations support this view.

The first observation is the microheterogeneity of the ensemble of
conformations that provide the same bent form of the double helix
during the last nanoseconds of both trajectories. Once the backbone
have found its preferred waving shape on the surface of the B-DNA
cylinder it fixes the bending direction. The thermal motion of bases
is allowed, but they are forced to respect this mechanical constraint,
giving rise to an ensemble of conformations with different base
orientations, but the same bent form of the double helix

The second observation is an always waving minor groove profile
which does not change during temporary short-living straightening.
The waving profile is the direct consequence of the postulated excess
of the specific backbone length over that in the regular B-DNA with
even grooves. The main immediate cause of bending is the necessity to
compress the backbone in the minor groove widenings if the parallel
base pair stacking is to be preserved \cite{Mzlanl:00a}.
The backbone stiffness tends to
cause destacking from the minor groove side, which results in bending
towards the major groove. Symmetrical destacking is also possible,
however, and transitions between various types of stacking
perturbations makes possible time variations of the magnitude of
bending with a constant backbone profile.

Finally, our model explains well the persistent bends in incorrect
directions and the rotation of the bending plane observed in TJA.
According to this view the excessive backbone length and its stiffness
force the backbone to wander on the surface of the B-DNA cylinder whatever
the base pair sequence is. In the dynamics starting from the A-DNA
structure the duplex enters the B-DNA family in strongly
non-equilibrium conditions, with rapidly changing different energy
components. The backbone quickly extends to its preferred specific
length taking some waving profile and causing bending in an arbitrary
direction. During the subsequent slow evolution it remains always
waving, and that is why there is always a preferred bend direction
which is not lost during occasional straightening.

\subsection*{Methods}

As in the previous report \cite{Mzlanl:00a}, the
molecular dynamics simulations were carried out with the internal
coordinate method (ICMD) \cite{Mzjcc:97,Mzjchp:99}. The
minimal model of B-DNA was used, with fixed all bond length,
rigid bases, fixed valence angles except those centered at sugar atoms
and increased effective inertia of hydrogen-only rigid bodies
as well as planar sugar angles \cite{Mzjacs:98,Mzlanl:00a}.
The time step was 10 fsec. AMBER94 \cite{AMBER94:,Cheatham:99}
force field and atom parameters were used with TIP3P water
\cite{TIP3P:} and no cut off schemes.
The heating and equilibration protocols were same as before
\cite{Mzjacs:98,Mzlanl:99}. The Berendsen algorithm
\cite{Berendsen:84} was applied during the production run,
with a relaxation time of 10 ps, to keep the temperature
close to 300 K. The coordinates were saved once in 2.5 ps.

The initial conformation for TJB was prepared by vacuum energy
minimization starting from the fiber B-DNA model constructed from the
published atom coordinates \cite{Arnott:72}. 375 water molecules were
next added by the hydration protocol designed to fill up the minor
groove \cite{Mzjacs:98}. The initial conformation for TJA
was prepared by hydrating the minor groove of the corresponding A-DNA
model \cite{Arnott:72} without preliminary energy minimization.
The necessary number of water molecules was added after
equilibration to make it equal in TJA and TJB.

During the runs, after every 200 ps, water
positions were checked in order to identify those penetrating into the
major groove and those completely separated. These molecules, if
found, were removed and next re-introduced in simulations by putting
them with zero velocities at random positions around the hydrated
duplex, so that they could readily re-join the core system.
This time interval was chosen so as to ensure a small enough average number
of repositioned molecules which was ca 1.5.

\section*{Appendix}
This section contains comments from anonymous referees of a peer-review
journal where this paper was been considered for publication,
but rejected (see also \onlinecite{Mzlanl:00a}).
\subsection{Journal of Molecular Biology}

\subsubsection {First referee}

These companion manuscripts describe a series of molecular dynamics
trajectories obtained for DNA sequences containing arrangements of oligo
dA - oligo dT motifs implicated in intrinsic DNA bending.  Unlike previous
MD studies of intrinsically bent DNA sequences, these calculations omit
explicit consideration of the role of counterions.  Because recent
crystallographic studies of A-tract-like DNA sequences have attributed
intrinsic bending to the localization of counterions in the minor groove, a
detailed understanding of the underlying basis of A-tract-dependent bending
and its relationship to DNA-counterion interactions would be an important
contribution.

Although the MD calculations seem to have been carried out with close
attention to detail, both manuscripts suffer from some troubling problems,
specifically:

The sequence investigated here is a 25-bp segment of the well-characterized
L. tarentolae kinetoplast-DNA bending locus.  Two trajectories, TJA and TJB,
were computed starting from canonical A-form and B-form structures,
respectively.  Although the author argues that greater structural
convergence between TJA and TJB has taken place in these simulations, there
is still a significant disparity concerning the observed bending directions
in these two structures.  Moreover, the extent of bending in this simulated
helix is significantly less than that observed in the previous study, which
is unexpected because of out-of-phase placement of the third A tract in the
previous sequence.  This behavior is not explained and seems difficult to
rationalize.

\end{document}